\documentclass[twocolumn,final,balancelastpage,superscriptaddress]{revtex4}

\usepackage{graphicx}

\usepackage{tabularx}
\usepackage{amssymb}
\usepackage{amsmath}
\usepackage{amsbsy}
\usepackage{comment}
\usepackage{mathrsfs}
\usepackage{dsfont}
\usepackage{nicefrac}
\usepackage{wrapfig}
\usepackage{amsfonts}
\usepackage{url} 
\usepackage[nolist]{acronym}
\usepackage{nicefrac}
\usepackage{subfigure}
\usepackage{epsf,color,colordvi,pifont}
\usepackage{showkeys}
\DeclareFontFamily{OT1}{pzc}{}
\DeclareFontShape{OT1}{pzc}{m}{it}{<-> s * [1.10] pzcmi7t}{}
\DeclareMathAlphabet{\mathpzc}{OT1}{pzc}{m}{it}

\begin{document}
\title{Light-amplified  Landau-Zener conductivity   in gapped graphene monolayers: \\a  simulacrum  of   photo-catalyzed vacuum instability}
\author{Selym \surname{Villalba-Ch\'avez}}
\email{villalba@uni-duesseldorf.de}
\author{Oliver \surname{Mathiak}}
\email{Oliver.Mathiak@hhu.de}
\author{Reinhold \surname{Egger}}
\email{egger@hhu.de}
\author{Carsten \surname{M\"{u}ller}}
\email{c.mueller@tp1.uni-duesseldorf.de }
\affiliation{Institut f\"{u}r Theoretische Physik, Heinrich-Heine-Universit\"{a}t D\"{u}sseldorf,\\ Universit\"{a}tsstr.\,1, 40225 D\"{u}sseldorf, Germany}

\begin{abstract}
Interband transitions of electrons in a gapped graphene monolayer are highly stimulated near the Fermi surface when a high-frequency electric wave of weak intensity and a strong constant electric field are superposed  in the plane of the  flake.  We consider this phenomenon equivalent to the Franz-Keldysh effect,  paying particular attention to the regime  where the photon energy linked to the fast-oscillating field is just below the graphene gap, so that the quantum transitions still occur through tunneling effects while being facilitated by the one-photon absorption channel.  In the considered parameter regime  the photo-catalyzed current linked to the described setup is shown to exceed   the one driven by the strong field solely by several orders of magnitude.  Conditions to relieve the impact of the field's finite extension are discussed, and a formula for the residual current density is derived.  The robustness of our assessment supports the viability of detecting this phenomenon in graphene,  thus providing a simulation  of the dynamically-assisted Schwinger mechanism in QED.
\end{abstract}

\date{\today}

\maketitle

\section{Introduction}

A strong, homogenous electric  background renders the ground state of quantum electrodynamics (QED) unstable, allowing quantum vacuum fluctuations of the electron-positron field to spontaneously materialize into their mass shells \cite{Gitman,Fradkin,Gitman1,Gitman2}. The pair production rate $\mathcal{R}\sim \exp[-\pi m_e^2c^3/(e\hbar E)]$  related to   this emblematic phenomenon---widely known  as the  Schwinger mechanism---has  a  tunneling nature and exhibits, in addition to a  nonlinear dependence on the external field $E$,  an essential singularity in the electric charge $e>0$    which transfers  to the process   a  nonperturbative feature \cite{Sauter:1931,Heisenberg:1935,Schwinger:1951nm}. So far, the observation of this field-induced vacuum instability has been prevented by the unavailability of field strengths comparable to the characteristic QED scale  $E_{\mathrm{cr}}=m_e^2 c^3/(e\hbar)\sim 10^{16}\ \rm V/cm$, at which the exponential suppression of  $\mathcal{R}$  turns out to be mitigated.\footnote{ Here  $m_e$  refers  to the electron mass and $c$ to the speed of light.  From now on  the Planck constant and the vacuum permitivity  are  set to unity $\hbar=\epsilon_0=1$.}   Although fields  of the order of   $E\sim10^{-2}E_{\mathrm{cr}}$   are   aimed for  by  the next generation of  multipetawatt laser  facilities such as the Extreme Light Infrastructure  (ELI) \cite{ELI}   and the Exawatt Center for Extreme Light Studies (XCELS) \cite{xcels}, it is generally believed that an  experimental realization of this yet hypothetical  vacuum breakdown  might represent  a major challenging task by the time when  both  ELI and XCELS become operational. The challenge is that  at the envisaged  field  strengths  the production rate remains very small, a fact that  calls   for  alternative routes which  relieve    the described issue.

While  in the last two decades theoretical  endeavors  toward  this goal  have  provided significant insight about  the nontrivial  nature of the quantum vacuum \cite{alkofer:2001ib,Hebenstreit:2009km,Bulanov,BlaschkeCPP,Kohlfurst:2012rb,Hebenstreit:2014lra,Gonoskov,Banerjee:2018gyt},   most notably  the  scenario first investigated  in Refs.~\cite{dgs2008, dgs2009}   has raised the hope of  observing  the  vacuum instability  for the first time.  In contrast  to the traditional  single-field  scheme,  the promising   setup   relies on  a temporal overlapping  of a  strong homogenous  electric field and a weak but high-frequency  $\omega\lesssim 2m_e$ electric pulse.  This  so-called  dynamically-assisted Schwinger mechanism retains  the  tunneling feature  while the   absorption of quanta   induces  a  reduction of    the  effective barrier width   that an electron has to traverse  from   the negative to the positive  Dirac continuum.  This reduction, in turn,  should facilitate the production of pairs  at  a   rate  $\mathcal{R}\sim \exp[-\kappa \pi E_{\rm cr}/E]$   that    enhances substantially  as compared to the case of the  standard Schwinger mechanism because the positive parameter $\kappa$ could be much smaller than unity.  Improvements  of various  field  configurations sharing the described idea  have been proposed,  including  the combination of a  static  electric field  and  pulses  with or without  subcycle structures \cite{Monin,Grobe2012,Linder:2015vta,Schneider,torgrimsson2016,Togrimsson1,Huang:2019uhf,Togrimsson2,Villalba-Chavez:2019jqp},  the situation in which  two Sauter waves are superposed   \cite{Orthaber2011,Fey},  as well as  scenarios in which both the strong and the fast-oscillating fields hold subcycles while being  modulated by  pulse profiles  \cite{Akal:2014eua,Otto:2014ssa,Otto:2015gla,Panferov:2015yda,Aleksandrov:2018uqb,Otto:2018jbs,Sitiwaldi:2018wad,taya}.  Furthermore, analogous upgrades  have been reported theoretically in both the Bethe-Heitler   \cite{DiPiazza:2009py,Augustin}  and   Breit-Wheeler   \cite{Jansen2013} pair production process.\footnote{In contrast to the  vacuum decay, there exists an  experimental confirmation of the nonlinear  Breit-Wheeler  pair production channel  in the few photon regime, where the rate follows a power law scaling with the applied field strength  \cite{Burke:1997ew}. New  campaigns  aim to verify  the fully   nonperturbative regime of this process   [see Refs.~\cite{Meuren2019,LUXE,LUXE2,Appleton,Salgado,Alina2022}]. } 

Despite the described advantage, gathering adequate  experimental conditions for  implementing  the  dynamically-assisted Schwinger mechanism   is not an easy task  because  strong fields  $E\sim E_{\mathrm{cr}}$ are required anyway for having a sizable production of pairs. Notwithstanding,  emergent condensed-matter systems with optoelectronic features that  resemble those linked to the QED vacuum  might  constitute solid state playgrounds  for simulating  the vacuum instability via  transitions of electrons   from valence to conduction  bands  of Landau-Zener nature \cite{Landau,Zener}.  However,  an  essential requirement  for assessing a plausible enhancement caused by the absorption of photons of  a weak but  fast-oscillating electric mode, i.e., the solid state analog of the dynamically-assisted Schwinger mechanism, is the existence of  \emph{massive} charge carriers.  A graphene monolayer \cite{GrExp1,GrExp2,GrExp3}  with a  tiny electronic band gap  $\Delta\sim1\; \rm meV$   acquired---for instance---by elastic strain engineering \cite{Strain,basov}, via  substrate-induced superlattices \cite{Substrate1,Substrate2}  or  through Rashba spin splittings on magnetic substrates \cite{Varykhalov},  is  perhaps the best  suited  platform for this purpose.  Mainly, because  the  charge carriers---with mass $\mathpzc{m}=\Delta/(2\mathpzc{v}_\mathrm{F}^2)$---in  this  two dimensional honeycomb lattice of  carbon atoms   possess a Diraclike dispersion relation near the neutrality  points  in which   the Fermi velocity $\mathpzc{v}_{\mathrm{F}}\approx c/300$ plays the role of the speed of light \cite{Wallace}. Hence, their behavior can be effectively described by a  $2+1$ dimensional Dirac model and their coupling to an electromagnetic field suitably simulates a planar QED with the particularization that  the creation of electron-hole pairs in band gapped graphene varieties  is predicted to occur at  a  rate $\mathcal{R}_{\; \mathpzc{g}}\sim \exp[-\pi E_{\mathpzc{g}}/E]$ that closely resembles the exponential dependence occurring in the Schwinger mechanism \cite{Mostepanenko,Akal:2016stu}.  However, in contrast to  $E_{\mathrm{cr}}$,  the  characteristic electric field  in graphene  with  $\Delta\approx 1\;\rm meV$,    $E_{\mathpzc{g}}=\Delta^2/(4e\mathpzc{v}_{\mathrm{F}})\approx 3.9\; \rm V/cm$    is rather easy to access  or even overpass,  opening in this way   an enticing  window  for  emulating the   dynamically-assisted vacuum breakdown   via Landau-Zener transitions catalyzed by the absorption of photons of a  fast-oscillating electric mode \cite{Akal:2018txb}.   It  is worth remarking  that   graphene flakes with tiny band gaps have  also  been  put forward  as toy environments to  test  intriguing low-dimensional effects in  the perturbative nonlinear regime of the  Breit-Wheeler-like  process   \cite{Golub:2019vzb,Golub:2021nhj}. 

Conceptually, the dynamically-assisted Landau-Zener effect must be understood as a relativistic-like generalization of the Franz-Keldysh effect occurring in semiconductors \cite{Franz, Keldysh}. So far, the microscopic study of this phenomenon has been carried out by adopting a nonrelativistic description based on the Schr\"odinger-band structure for holes and electrons. Its main phenomenological aspect is the modification of the bulk optical properties brought about by a strong, slowly varying electric field.   These deviations become noticeable in the quasiparticle spectrum, which turns out to be finite even below the bandgap and oscillatory when the energy of the absorbed photon lies above it. The described modifications are immediate consequences of the nonperturbative interplay between the valence and conduction bands and the strong electric field background, all together playing the role of the polarized QED vacuum. While the Franz-Keldysh effect has been experimentally verified in  three-dimensional ordinary semiconductors \cite{RWilliams,HIRalph,Jauho,Nordstrom,Putnam}, we are aware neither of theoretical studies nor experimental investigations aiming to determine its inherent electronic properties when the material  has low dimensionality  and its  description admits a Dirac model, i.e., where holes and electrons are not independent. With its exceptional conductive properties, graphene provides, perhaps, the most direct access to this scenario \cite{Berdyugin,ASchmitt}. From this perspective, the study of the Franz-Keldysh phenomenon in gapped graphene   is worthwhile on its own,  offering in this way  a practical method for injecting free-of-contact ultrafast currents with subcritical fields and for controlling the conductivity of the material by adjusting the fast-oscillating laser wave's parameters.
 
We should stress  at this point the substantial   efforts  devoted to  simulate  various relativistic  processes   in graphene layers  with   gapless band structures  [$\Delta=0$].     See  for instance  Refs.~\cite{Neto,kotov} and references therein. Indeed, the first theoretical Landau-Zener studies in graphene were carried out by considering  massless electronic excitations \cite{Allor:2007ei,Mostepanenko,Lewkowiczprl,dora1,Lewkowiczprb,Avetissian,Fillion-Gourdeau:2015dga,Fillion-Gourdeau:2016izx}. Subsequent   measurements of   optical radiation---emitted  presumably  by the recombinations of  residual  electron-hole pairs  produced  via the aforementioned mechanism  \cite{Oladyshkin} [see also Refs.~\cite{Lui,Tani}]---and  currents induced by two-cycle laser pulses  \cite{Takuya,Heide} confirmed the phenomenon. Besides, the detection of the current  turned out to be sensitive to the carrier envelope phase of the driving field which facilitates  a coherent control of  massless electron dynamics and  emulates a  prediction expected within  the Schwinger mechanism  \cite{Hebenstreit:2009km}.  Clearly, similar  setups   could  be implemented to probe   field-induced interband transitions of \emph{massive} quasiparticles  stimulated by the absorption of photons  of a weak but fast-oscillating wave.  As  in this scenario  the transition rate is enhanced,   the yielded electron-hole pairs would  have  a density higher than  in the absence of the weak field, provided    $E_{\mathpzc{g}}\gg E$.   It is then likely that the recombination and thus the emission of photons become noticeable, or alternatively, that a sizable current can be measured to verify the solid-state analog of the dynamically-assisted Schwinger mechanism.  This paper is devoted to theoretically investigating the latter possibility.   We propose a realistic graphene-based setup to examine the analog of the dynamically assisted Schwinger effect. Unlike earlier studies \cite{Linder,Akal:2018txb}, which mostly dealt with the associated transition rate, our emphasis is primarily focused on the field-induced current caused by the transition of electrons from valence to conduction band, and the impact of the field's finite-size on this observable.  Particularly,  we reveal that the detection of the process benefits when the field's on and  off switching  occurs  through smooth ramping and deramping sectors. Our effort is oriented toward optimizing the amplification that the current undergoes when a strong electric field is active, and quanta from a fast-oscillating wave are absorbed in the meantime. 

This paper is organized as follows.  In Sec.~\ref{II},  we describe the model to be analyzed and  briefly summarize  the main aspects linked to  the  quantum kinetic equation to be used, addressing in this way  similarities  with the pair production process in QED  \cite{Schmidt:1998vi,Kluger,Schmidt:1999vi}.   Properties of the quasiparticle spectrum are elucidated in Sec.~\ref{SPDFasymptotic}. Particular attention is paid to the impact on the quasiparticle spectrum of strong backgrounds characterized by abrupt versus smooth turn-on/off sectors.  A compact asymptotic formula for the current density related to the residual number of excitations is  derived in Sec.~\ref{observabledcp}.   Later on,   we compare numerical and analytical predictions and identify the parameters that ensure an optimal current due to the photo-catalyzation of Landau-Zener transitions. We conclude the  paper  in Sec.~\ref{sec:numerical} with  an overview of our main results.

\section{General aspects\label{II}}

\begin{figure}
\includegraphics[width=0.50\textwidth]{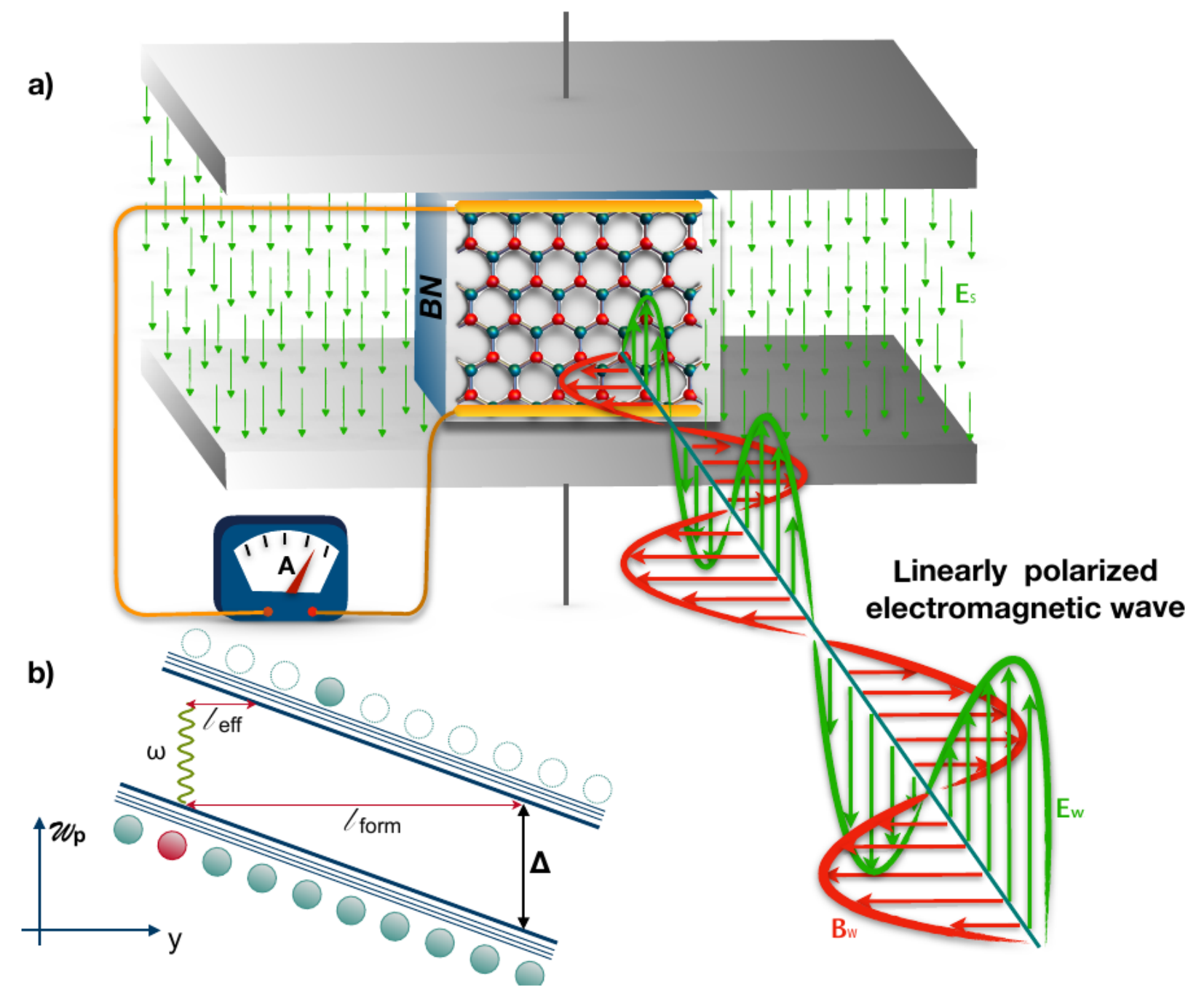}
\caption{a) Scheme of   an experimental set-up allowing the  simulation of the dynamically-assisted Schwinger mechanism in band-gapped graphene via the production of electron-hole  pairs.  A  graphene flake---grown  on top of a substrate---is placed  within the plates of a capacitor which holds a strong electric field $E_s$. Simultaneously,  the graphene sheet is  irradiated with  a  linearly polarized plane-wave with  frequency $\omega$, the amplitude of which $E_w$ is supposed to be weaker than the field generated by the capacitor [$E_w\ll E_s$]. The field-induced current  in the graphene flake is then measured with the help of  an ammeter. To this end,  two electrodes are deposited on  the stripe.  b) The tilt that the relativisticlike dispersion relation undergoes owing  to the strong electric field makes it possible for electrons---blobs colored in cyan---to tunnel from the valence band to the initially empty  conduction band. The absorption of a photon of energy slightly below the band gap [$\omega<\Delta$]  reduces the effective distance needed to reach the latter.}
\label{fig:1}
\end{figure}

Let us  consider the spontaneous production of electron-hole pairs taking place in a time-dependent but homogeneous electric field combining a strong static mode with  strength $E_s$ 
and a perturbative monochromatic wave with amplitude  $E_w$   [$E_w\ll E_s$]  and frequency $\omega$. Hereafter we will assume both fields  localized temporally between $-T/2\leqslant t\leqslant T/2$, 
so that the pulse length of the wave  $T=2\pi N /\omega$   can be written  in terms of the number of cycles $N$.  The corresponding  four-potential reads
\begin{equation}
\begin{split}
\mathpzc{A}^\mu(t)&=\flat^{\mu}\left\{\begin{array}{cc}\\ -cE_s t-\frac{c E_w}{\omega}\sin(\omega t),&\vert t\vert \leqslant\frac{1}{2}T\\\\ 0,& \mathrm{otherwise}\end{array}\right.
\end{split}\label{EField}
\end{equation}where $\flat^\mu$ is the polarization  four-vector.  In practice, the  strong field  might result from  a capacitor with a dc voltage in which the graphene sheet is placed  [see Fig.~\ref{fig:1}a].  Conversely, a fast-oscillating electric wave can be successfully generated from an incident laser beam with a polarization parallel to the strong field direction.  Observe that, to fit with our theoretical treatment, the waist size $w_0$ of the linearly polarized laser wave has to be  much larger than the length of the graphene surface, which we take  here of the order of $\ell\gtrsim100\;\rm \mu m$. 
The described mechanism does not provide a pure  electric wave as it is required by Eq.~(\ref{EField}). Indeed, the incident  wave---which can be thought as a plane-wave---is not homogeneous and, thus, a magnetic field parallel to the graphene surface would be present. However, the two-dimensional confinement of the quasiparticles  prevents any influence of this field component on their dynamics. Hence,  what  they  undergo actually  is  nothing but the combination of  a weak electric field oscillating in time and the strong  field linked to the capacitor.\footnote{Let us briefly mention that  the source of the strong field could alternatively be generated from an additional laser beam with a frequency   $\Omega\ll \omega$ and intensity higher than those associated with the fast-oscillating wave. The upcoming study must be then understood in such a situation  as the leading order contribution of an adiabatic approximation. }  Clearly, this will be the case  whenever the used graphene monolayer is perfectly plane-shaped. In practice, however, free-standing flakes are rippled and contain distortions. These problems can be  mitigated to a large extent if the graphene sample is grown on top of a  hBN substrate, enabling one to obtain an almost ideally flat surface while simultaneously inducing the required band gap $\Delta$.  To suitably simulate the dynamically assisted Schwinger mechanism, this gap has to simultaneously satisfy two  conditions:  $\Delta^2\gg4 eE_s \mathpzc{v}_{\mathrm{F}}$  and  $\Delta\gtrsim\omega$.  For a band gap $\Delta\gg \omega$, the effective reduction of the barrier width   [see Fig.~\ref{fig:1}b] between the valence  and conducting bands  $\mathpzc{l}_{\mathrm{eff}}\sim(\Delta-\omega)/(eE_s)$ approximates to the  formation length $\mathpzc{l}_{\mathrm{form}}\sim\Delta/(eE_s)$ associated with  the production of electron-hole pairs  when the strong field  is present only. We remark that the generation of the gap  $\Delta$ through  a substrate-induced mechanism  implies that, because of the hBN dielectric constant   $\varepsilon_{\mathrm{hBN}}\approx 3.4$ \cite{pierret}, the field strength that the graphene sheet experiences is diminished by a factor of $2/(1+\varepsilon_{\mathrm{hBN}})\approx0.4$ when compared to the vacuum situation. Consequently, both $E_s$ and $E_w$ are supposed to take  this reduction into account.

Here,  the  inter-band transitions of electrons will be  investigated by adopting  a quantum  kinetic approach. This formulation---which is equivalent to other well-known approaches based on  QED in unstable  vacuum \cite{Gitman,Fradkin,Gitman1,Gitman2}---comprises the  dynamical information of the  pair production  process in the single-quasiparticle distribution function  $W_{\mathpzc{g}}(\pmb{p};t)$,  which refers  to  fixed spin and valley quantum numbers and  relaxes to those linked to the electrons and holes  only when the total external field  is switched off $\pmb{E}(\pm\infty)\to 0$, i.e. formally  at $t\to\pm\infty$. In this context, the quantum Boltzmann-Vlasov equation which dictates the time evolution of 
$W_{\mathpzc{g}}(\pmb{p};t)$ reads \cite{Akal:2016stu,Akal:2018txb,Smolyanskygraphene,Gitmangraphene}:
\begin{eqnarray}\label{vlasovgraphene}
&&\dot{W}_{\mathpzc{g}}(\pmb{p};t)=Q(\pmb{p},t)\int_{-\infty}^t d\tilde{t}\; Q(\pmb{p},\tilde{t})
\left[\frac{1}{2}-W_{\mathpzc{g}}(\pmb{p};\tilde{t})\right]\nonumber\\&&\qquad\quad\qquad\times\cos\left[2\int_{\tilde{t}}^t dt^{\prime}\ \mathpzc{w}_{\pmb{p}}(t^{\prime})\right],
\end{eqnarray}where  the   initial condition $W_{\mathpzc{g}}(\pmb{p},-\infty)=0$, i.e., an initially empty conduction band is  assumed. This formula is characterized by the function $Q(\pmb{p},t)\equiv e E(t)\mathpzc{v}_{\mathrm{F}} \epsilon_\perp /\mathpzc{w}_{\pmb{p}}^2(t)$,   which depends on the quantity  $\epsilon_\perp=[\frac{1}{4}\Delta^2+\pi_\perp^{2}\mathpzc{v}_{\mathrm{F}}^2]^{1/2}$  and the respective  total energy  $\mathpzc{w}_{\pmb{p}}(t)=[\epsilon_\perp^2+\pi_\parallel^2(t)\mathpzc{v}_{\mathrm{F}}^2]^{1/2}$.  Hereafter, $\pi_\perp=p_\perp$ and $\pi_\parallel(t)=p_\parallel-e\mathpzc{A}(t)/c$  will refer to the components of  kinetic  momentum  $\pmb{\pi}(t)=(\pi_\perp,\pi_\parallel)$ of the quasiparticle  perpendicular and parallel to the direction of the electric field $\pmb{E}(t)=-\partial \pmb{\mathpzc{A}}/\partial(ct)=\pmb{E}_s+\pmb{E}_w\cos(\omega t)$ [$\mathpzc{A}_0(t)=0$], respectively.   At this point,  it is worth noting that  $\pmb{\pi}(t)$ has to be understood relative  to either  $\pmb{K}$  or $\pmb{K^\prime}$ points,  satisfying the condition $\vert\pmb{\pi}(t)\vert\ll\vert\pmb{K}^{(\prime)}\vert\approx 3 \ \mathrm{eV}/\mathpzc{v}_{\mathrm{F}}$.  
To avoid the electric field causing a shift in the quasiparticle momentum $\Delta \pi=eE_sT$  comparable to $\vert\pmb{K}^{(\prime)}\vert$ and thus to prevent the wave packet from crossing the Brillouin zone boundary, we shall consider field operating times   $T$ much below the time scale  \cite{dora1,RosensteinRC}
\begin{equation}\label{Blochscale}
T_{\mathrm{Bloch}}= \frac{\vert\pmb{K}\vert}{eE_s}
\end{equation}
on which the Bloch oscillations come into play. On the other hand,  $T$ is expected to be much longer than the characteristic formation time of an electron-hole pair, i.e. $\mathpzc{l}_{\mathrm{eff}}/\mathpzc{v}_{\mathrm{F}}$ with  $\mathpzc{l}_\mathrm{eff}\sim(\Delta-\omega)/(eE_s)$  denoting the effective reduction of the barrier width between the valence  and conduction bands  [see Fig.~\ref{fig:1}b]. The constraints on the operating time can be summarized by the following relation:
\begin{equation}\label{timerestrictions}
\frac{\Delta}{eE_s\mathpzc{v}_\mathrm{F}}\ll T\ll  \frac{\vert\pmb{K}\vert}{eE_s}.
\end{equation}We remark that,  as no dependence on temperature is manifested in Eq.~(\ref{vlasovgraphene}),  any outcome resulting from it must be  interpreted within the zero temperature limit. Therefore, at asymptotically earlier times $t\to-\infty$ for which  $E(t)$ vanishes, the system  behaves  as a  degenerate  Fermi gas. In line with this limit, the  quasiparticle energy  $\mathpzc{w}_{\pmb{p}}(-\infty)=[\pmb{p}^2\mathpzc{v}_{\mathrm{F}}^2+\frac{1}{4}\Delta^2]^{1/2}$   has to be understood relative to the Fermi-level, here assumed at $\varepsilon_{\rm F}=0$.  We emphasize that  backreaction effects \cite{Bloch1999} are neglected throughout the paper.

There is a tight parallelism between the process under consideration and the spontaneous production of pairs from the vacuum polarized by a similar background field setup [see Eq.~\eqref{EField}]. It is therefore not surprising that, after some modifications, corresponding QED findings can be directly applied to the interband transitions of electrons in a gapped graphene monolayer. In this sense, we should mention that an analytic expression for the single-particle distribution function in the dynamically-assisted field configuration has been derived in Ref.~\cite{taya}.  For its establishment, the author carried  out a calculation in which the interaction caused by the strong field was considered through the retarded Green's function method without any approximation, whereas the interplay with the fast oscillating wave was handled perturbatively. Such a  formula can be readily adapted to the graphene scenario  by applying the recipe given in Ref.~\cite{Akal:2016stu}   and reads: 
\begin{equation}
\begin{split}
W_{\mathpzc{g}}(\pmb{p})&\approx e^{-\frac{\pi\epsilon_\perp^2}{eE_s\mathpzc{v}_{\mathrm{F}}}}\left\vert1+\varepsilon\frac{\pi}{2}\frac{\epsilon_\perp^2}{eE_s\mathpzc{v}_{\mathrm{F}}}e^{-\frac{i}{2}\gamma\left(\frac{\omega}{\Delta}+4\frac{p_\parallel\mathpzc{v}_\mathrm{F}}{\Delta}\right)}\right.\\
&\qquad\qquad\times \left. _1\tilde{F}_1\left(1-\frac{i}{2}\frac{\epsilon_\perp^2}{eE_s\mathpzc{v}_{\mathrm{F}}};2;i\gamma \frac{\omega}{\Delta}\right)\right\vert^2.
\end{split}
\label{distributionfunction}
\end{equation}Here,  $\varepsilon=E_w/E_s$ parametrizes the relative weakness of the fast-oscillating  mode,  $_1\tilde{F}_1(a;b;z)$ is the regularized hypergeometric function \cite{NIST}, and $\gamma=\omega \Delta/(2eE_s\mathpzc{v}_{\mathrm{F}})$  denotes the combined Keldysh parameter.    Noteworthy, no matter how fast or slow the oscillating mode is, Eq.~\eqref{distributionfunction} applies whenever its amplitude is significantly weaker than the strong field strength, i.e. $E_w\ll E_s$.  Besides, there is no restriction on the frequency of the fast-oscillating wave; it can be either below or above the energy gap. The important point here is that  it is valid whenever the interaction time is  larger than any characteristic time scale linked to  the pair production process, including $2 \vert p_{\perp,\parallel}\vert/(eE_s)\ll T$.  The field-stimulated transitions of electrons from valence to conduction bands can  be studied using the formula above if $T$, in addition, satisfies  the restriction given in Eq.~\eqref{timerestrictions}.

\begin{figure*}
\centering
\includegraphics[width=1\textwidth]{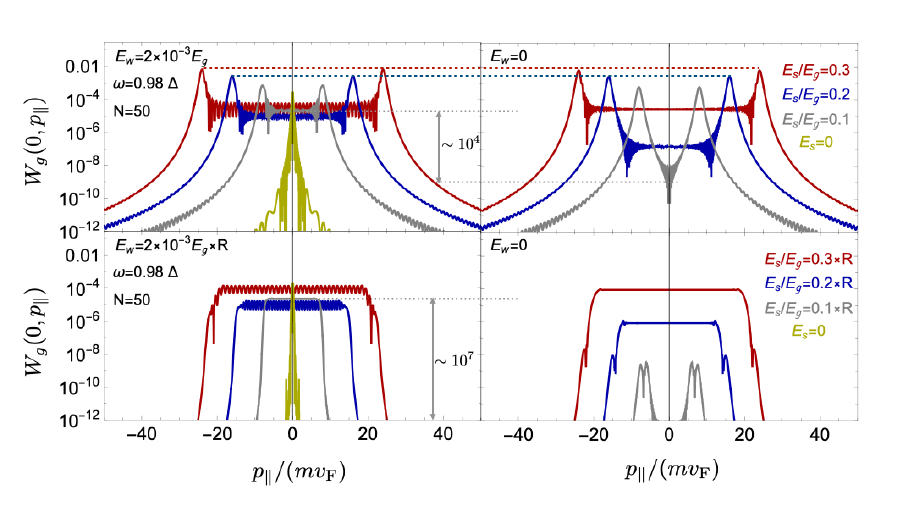}\vspace{-.5cm}
\caption{Long-time behavior of the single-quasiparticle distribution function.  The dependence of the spectrum on  the momentum parallel  $p_\parallel$  to the external electric field at $p_\perp=0$  and without ramping and deramping sectors (i.e. sudden tun on and off)  is depicted in the upper panel. The lower panel shows the results associated with the scenario that includes smooth turn-on and turn-off phases. The electric amplitude of the rapidly oscillating wave $E_w = 2\times 10^{-3} E_{\mathpzc{g}}$   was used to generate the left upper pictures. In contrast, the one on its right, which serves as a comparison, considers the situation in which the weak mode is turned off.  The fields used in the lower panel have been adjusted with the parameter $R=1.12$.  Curves sharing  the  same  color  have  been obtained by using the same parameters. The numerical outcomes have been obtained by setting the number of cycles $N=50$, which corresponds  to a time period  $T=30\; \rm ps$ if the band gap is chosen as $\Delta=7\; \rm meV$ and $\omega=0.98\Delta$. The latter was adopted to prevent Bloch's oscillations from being caused by the field. The corresponding scales [see Eq.~\eqref{Blochscale}] are $T_{\mathrm{Bloch}}=0.54\;\rm ns$  for $E_s=0.3\; E_{\mathpzc{g}}$, $T_{\mathrm{Bloch}}=0.82\;\rm ns$   for $E_s=0.2\; E_{\mathpzc{g}}$ and $T_{\mathrm{Bloch}}=1.6\;\rm ns$  for  $E_s=0.1\; E_{\mathpzc{g}}$. Here, the critical field strength  is $E_{\mathpzc{g}}=\Delta^2/(4e\mathpzc{v}_{\mathrm{F}})=0.19\; \rm kV/cm$, whereas   the mass of the carriers is  $\mathpzc{m}=\Delta/(2\mathpzc{v}_\mathrm{F}^2)=0.32\; \mathrm{keV}/c^2$.}
\label{fig:1.1}
\end{figure*}

We note that, for $\gamma\ll  \Delta/\omega$, i.e. for small frequencies $\omega\ll \sqrt{2eE_s\mathpzc{v}_{\mathrm{F}}}$,  Eq.~\eqref{distributionfunction} reproduces the Landau-Zener formula
\begin{equation}
W_{\mathpzc{g}}(\pmb{p})\approx e^{-\frac{\pi\epsilon_\perp^2}{eE_s\mathpzc{v}_{\mathrm{F}}}}\left\vert1+\frac{\varepsilon \pi}{2}\frac{\epsilon^2_\perp}{eE_s}\right\vert^2
\label{LZcorrections}\end{equation}up to $\mathcal{O}(\varepsilon^2)$. In the opposite regime $\gamma\gg  \Delta/\omega$, i.e., for high frequencies $\omega\gg \sqrt{2eE_s\mathpzc{v}_{\mathrm{F}}}$, Eq.~\eqref{distributionfunction}  combines both perturbative $\sim\varepsilon$ and nonperturbative $\sim\exp[-\pi\epsilon_\perp^2/(eE_s\mathpzc{v}_{\mathrm{F}})]$ contributions. However, following the asymptotic study made in Ref.~\cite{taya}, if   $eE_s\mathpzc{v}_{\mathrm{F}}\ll \epsilon_\perp^2$, the absorptive perturbative contribution dominates and  
\begin{equation}
W_{\mathpzc{g}}(\pmb{p})\approx T\varepsilon^2 \frac{\pi}{8} \frac{\epsilon_\perp^2\mathpzc{v}_{\mathrm{F}}^2}{\mathpzc{w}_{\pmb{p}}^2(-\infty)}\frac{(eE_s)^2}{\mathpzc{w}_{\pmb{p}}^2 (-\infty)}\delta(\omega-2\mathpzc{w}_{\pmb{p}}(-\infty)),
\label{largekeldych}
\end{equation}where $\mathpzc{w}_{\pmb{p}}(-\infty)=[\pmb{p}^2\mathpzc{v}_{\mathrm{F}}^2+\frac{1}{4}\Delta^2]^{1/2}$ is the free dispersion relation of the quasiparticles. We note that the estimated solution to the transport equation [see Eq.~\eqref{vlasovgraphene}]  that results when the low-density approximation $1\gg2 W_{\mathpzc{g}}(\pmb{p};t)$ is used, makes it simple to verify both Eqs.~\eqref{LZcorrections} and \eqref{largekeldych}.

\section{Properties of  the quasiparticle spectrum\label{SPDFasymptotic}}

In this section, we will address how to enhance the likelihood of observing dynamically assisted Landau-Zener transitions in the proposed setting. A first step toward this aim is to understand how the solution to  Eq.~\eqref{vlasovgraphene}  responds to changes in the parameter space. However, to deal with numerical evaluations, an equivalent system of ordinary differential equations rather than the integro-differential form of the quantum Boltzmann-Vlasov equation will be used:
\begin{equation}\label{firstequa}
\begin{split}
&i\dot{f}(\pmb{p},t)=\mathpzc{a}_{\pmb{p}}(t)f(\pmb{p},t)+\mathpzc{b}_{\pmb{p}}(t)g(\pmb{p},t),\\ 
&i\dot{g}(\pmb{p},t)=\mathpzc{b}^*_{\pmb{p}}(t)f(\pmb{p},t)-\mathpzc{a}_{\pmb{p}}(t)g(\pmb{p},t).
\end{split}
\end{equation}Here $f(\pmb{p},t)$ and $g(\pmb{p},t)$ are Bogoliubov coefficients \cite{Akal:2014eua,Akal:2016stu,Mostepanenko:1972,Bagrov,Mostepanenko} and  $W_{\mathpzc{g}}(\pmb{p};t)=\vert f(\pmb{p},t)\vert^2$. The initial conditions $g(\pmb{p},-T/2)=1$ and $f(\pmb{p},-T/2)=0$ are  then chosen.  The remaining  elements contained in the equations above   are 
\begin{eqnarray}\label{coefficient1}
\begin{array}{c}
\displaystyle\mathpzc{a}_{\pmb{p}}(t)=\mathpzc{w}_{\pmb{p}}(t)+\frac{eE(t)p_\perp\mathpzc{v}_\mathrm{F}^2}{2\mathpzc{w}_{\pmb{p}}(t)\left(\mathpzc{w}_{\pmb{p}}(t)+\frac{1}{2}\Delta\right)},\\ \\
\displaystyle\mathpzc{b}_{\pmb{p}}(t)=\frac{1}{2}\frac{eE(t)\epsilon_\perp}{\mathpzc{w}_{\pmb{p}}^2(t)}\exp\left[-i\tan^{-1}\left(\frac{p_\perp \pi_\parallel\mathpzc{v}_\mathrm{F}^2}{\epsilon_\perp^2+\frac{1}{2}\mathpzc{w}_{\pmb{p}}(t)\Delta}\right)\right].
\end{array}
\label{coefficient2}
\end{eqnarray}  By changing the momentum components within the range $-0.18\;  \mathrm{eV}\leqslant p_{\perp,\parallel}\mathpzc{v}_\mathrm{\mathrm{F}}\leqslant  0.18\;  \mathrm{eV}$, the system of differential equations described above have been solved. In this assessment, a  band gap  $\Delta=7\; \rm meV$ and frequency $\omega=0.98\Delta$ were utilized.   Separately, three distinct strong field strengths $E_s=(0.1, 0.2, 0.3)E_{\mathpzc{g}}$  were applied,   with $E_{\mathpzc{g}}=\Delta^2/(4e\mathpzc{v}_{\mathrm{F}})\approx 0.19\; \rm kV/cm$  denoting the critical electric field strength. Besides,  the number of cycles of the fast-oscillating wave was chosen as $N=50$ to guarantee  that the field operating time  $T=30\; \rm ps$  is much shorter than the minimum Bloch scale $T_{\mathrm{Bloch}}=0.54\;\rm ns$, corresponding to $E_s=0.3 E_{\mathpzc{g}}$. 

\begin{figure}
\centering
\includegraphics[width=3.2in]{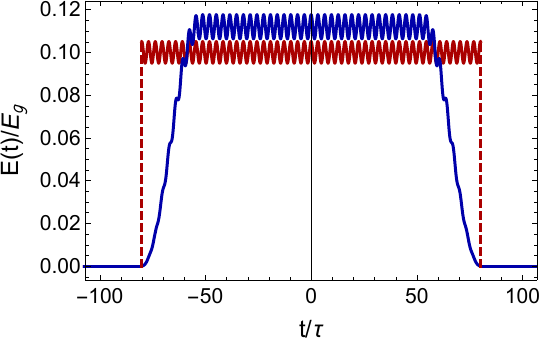}
\caption{Time-dependence of the electric field with abrupt turn-on/off sectors (in red) and smooth ramping and deramping phases with $\sin^2[\frac{1}{2}\pi(1-6t/T)]$ shape (in blue). To ensure that the energy of the two pulses is equal, the curve associated with the latter scenario has been multiplied by the factor $R=1.12$. Here, the time is given in units of $\tau=\Delta^{-1}$ and $E_w=5\times 10^{-3}E_{\mathpzc{g}}$. }
\label{fig:2v}
\end{figure}

The results of this investigation are displayed in the upper panel of Fig.~\ref{fig:1.1}. Both pictures  show  the long-term dependence  of the quasiparticle spectrum  
\begin{equation} 
W_{\mathpzc{g}}(\pmb{p})\equiv \lim_{t\to T/2}W_{\mathpzc{g}}(\pmb{p};t)
\end{equation}  on the longitudinal momentum $p_\parallel$ of the produced excitation at $p_\perp=0$.  The upper   left  picture takes into account a rapidly oscillating mode with strength $E_w=2\times 10^{-3}E_{\mathpzc{g}}$, whereas the one on its right considers solely the influence of the strong field [$E_w=0$]. In both panels, the  solid curves exhibit  two maxima symmetrically spaced from $p_\parallel=0$.  Their positions $p^{\mathrm{max}}_\parallel=\pm\frac{1}{2} eE_sT$ match up with the quasiparticle vanishing kinetic momentum [$\pi_\parallel(\pm T/2)=0$].  We note that the electron and hole distribution functions share the same shape because the charge conjugation, space inversion, and time reversal symmetries are naturally included in the Dirac model. Hence, if an electron is promoted to the conduction band in a state characterized by any of the peaks exhibited in Fig.~\ref{fig:1.1}, a hole will likewise be produced at the peak of its spectrum, where its momentum is opposite to that of the electron. This indicates that the creation of pairs is most likely to occur with their constituents  occupying low-energy states $2\mathpzc{w}_{\;\pmb{p}}(\pm T/2)=\Delta$. For a specific strong field strength, a comparison of the left and right upper pictures in  Fig.~\ref{fig:1.1}  renders it readily evident that neither the peak heights nor their widths are influenced by the assisting  fast-oscillating mode. Therefore, the origin and phenomenology linked to these maxima are attributable to the finite extension of the strong field only. Indeed, both peaks  result from the sharp behavior of the electric field at its edges.

\begin{figure*}
\centering
\includegraphics[width=1\textwidth]{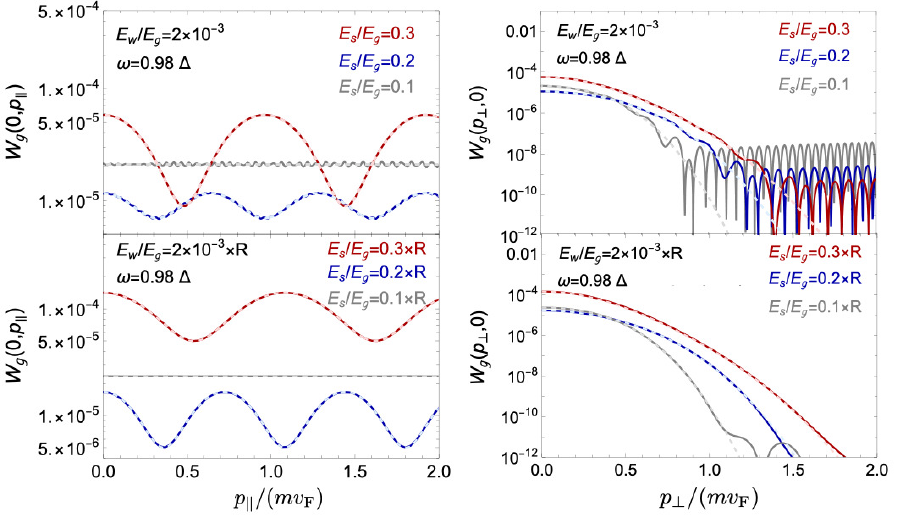}
\caption{Comparison between the numerical and analytical  dependence of $W_{\mathpzc{g}}$ on the  momentum parallel to the external electric field $p_\parallel$  at $p_\perp=0$ (left panel).  The  dependence on the momentum perpendicular to the external electric field at $p_\parallel=0$ is shown on the right. The  results in the top band were obtained by considering a background with sharp edges, while those in the lower band included smooth on/off sectors. The  behavior resulting  from the analytical formula [see Eq.~\eqref{distributionfunction}] is depicted in dashed light color style.  Curves sharing  the  same  color  have  been obtained by using the same parameters. The numerical outcomes have been obtained by setting the number of cycles $N=200$, which corresponds  to a time  $T=0.12\; \rm ns$ if the band gap is chosen as $\Delta=7\; \rm meV$.} 
\label{fig:2}
\end{figure*}

Experimentally, the field can be continuously varied from a situation in which the turn-on is very quick---almost abrupt---to one in which the turn-on is slow and smooth. In such a situation, the height of the peaks drops and may even vanish completely. To assure the latter scenario, the time interval needed to turn the field on or off must be longer than the characteristic formation period for an electron-hole pair. Otherwise, the electron is promoted almost instantly to the conduction band without having changed practically its initial state [see below Eq.~\eqref{timerestrictions}]. Observe that this viewpoint enables us to interpret the peak on the right (left) of the spectrum as being due to the abrupt turn-on (off). The lower panel of Fig.~\ref{fig:1.1}  is meant to support this statement.  There we assess the impact of incorporating  turning on and turning off  field phases on the quasiparticle spectrum.  For this, the external field was modulated with a ramping (deramping) function with a $\sin^2[\frac{1}{2}\pi(1-6t/T)]$-shape during a sixth of the total interacting time $T$, while for the remaining $2T/3$-period  it is driven by the field  $E(t)=E_s+E_w\cos(\omega t)$. To make a fair comparison, we have adjusted this modified background such that the energy delivered to the graphene sample is the same as the one provided  by the  background in Eq.~\eqref{EField}. The correction is implemented by multiplying the field and related  potential by a constant coefficient  $R\approx\sqrt{24/19}=1.12$. Figure~\ref{fig:2v} depicts the field's shape both with and without the ramping and deramping components. It is worth mentioning that the $\sin^2-$functions used in the implementation of the  latter do not imply that the corresponding gauge potential is smoothly  switched  on and off  when $t\to \pm T/2$. Indeed, throughout the calculations, this one has been taken continuously within the interacting time $T$, but with sharp turn-on/off sectors. We remark that the $\sin^2$ choice yields a distribution function that is devoid of the  maxima previously discussed while retaining the oscillatory  sectors  that are in between them.\footnote{We have checked that the discontinuous behavior of  the potential $\mathpzc{A}(t)$ at $t=\pm T/2$ does not affect the outcome of our numerical calculations.}

Comparing the left and right panels in Fig.~\ref{fig:1.1}, one notices that the exhibited $p_\parallel$-region is  indeed sensitive to the fast-oscillating mode. However, the modifications will be noticeable depending on how strong $E_s$ is.  For instance, at $E_s=0.1 E_{\mathpzc{g}}$ (curves in light  gray) and $p_\parallel=0$, the value on the left lower panel exceeds by seven  orders of magnitude the one on its right. Yet, if $p_\parallel\in[5,10]\mathpzc{mv}_{\mathrm{F}}$, for  example, this difference is smaller due to  contiguous peaks that the spectrum without the fast-oscillating wave  exhibits. The appearance of these maxima--originated by side-band terms due to the ramping (deramping) functions--contrasts with the known Landau-Zener formula  $W_{\mathpzc{g}}\approx \exp[-\pi\epsilon_\perp^2/(eE_s\mathpzc{v}_{\mathrm{F}})]$, which is independent of the parallel momentum of the yielded quasiparticles. This fact  provides another evidence that  the distribution function $W_{\mathpzc{g}}$, and thus, its associated observables, are sensitive to the field profile.  We note, however, that as the interaction time grows, the height of these side-band-induced peaks gradually decreases.  Under such circumstances, it could be difficult to visualize their impact.

A distinctive property of the spectrum assisted dynamically [see lower left  picture in Fig.~\ref{fig:1.1}] is that its highest portion oscillates along $p_\parallel$, and that the amplitudes of these oscillations diminish as the strong field weakens. 
Besides,  the findings in the left panel  display a non-monotonic trend with the decreasing strength of the strong field. Indeed,  for $E_s = 0.1 E_{\mathpzc{g}}$ the associated spectrum's oscillatory  sector exceeds the one linked to $E_s = 0.2 E_{\mathpzc{g}}$. This behavior does not occur in the right panel and can thus be attributed to the presence of the fast oscillating wave.  All these features are clearly illustrated in Fig.~\ref{fig:2}.  Its picture at the top corresponds to the case in which the field has abrupt on/off edges, while the one at the bottom comprises smooth functions in these initial and final phases of the field configuration. The corresponding outcomes resulting from the analytic formula given in  Eq.~\eqref{distributionfunction} are included. They are exhibited in dashed light color style, whereas the numerical outcomes are shown by solid lines. Indeed, the numerical and approximate curves obtained at a strong field strength share the same color, although with a different color intensity.  Almost no difference is discernible between the exact outcomes and those resulting from Eq.~\eqref{distributionfunction}; for each strong electric field, both curves overlap one on top of the other. This analysis indicates that the ridges of the oscillatory sector are reached  at  $p_\parallel=eE_s \tau_{\mathrm{rid}}$ with $ \tau_{\mathrm{rid}}=2\pi \mathpzc{K}/\omega$, $\mathpzc{K}\in\mathbb{Z}$ and $\vert\mathpzc{K}\vert<\frac{1}{2}N$. Besides, the region containing these maxima extends over $\Delta p_\parallel\approx \frac{2}{3}eE_s T$, which indicates that the yielded pair  is most likely  to be localized  within a spatial interval  $\Delta x_\parallel \gtrsim 1/\Delta p_\parallel=(\frac{2}{3}eE_sT)^{-1}=180\;\rm  \AA$ if $E_s=0.1E_{\mathpzc{g}}$. To put the uncertainty  into perspective, it is   seventy  times  greater than the  lattice constant of graphene  [$a_0 = 2.46\; \rm\AA$].

\begin{figure*}
\centering
\includegraphics[width=1\textwidth]{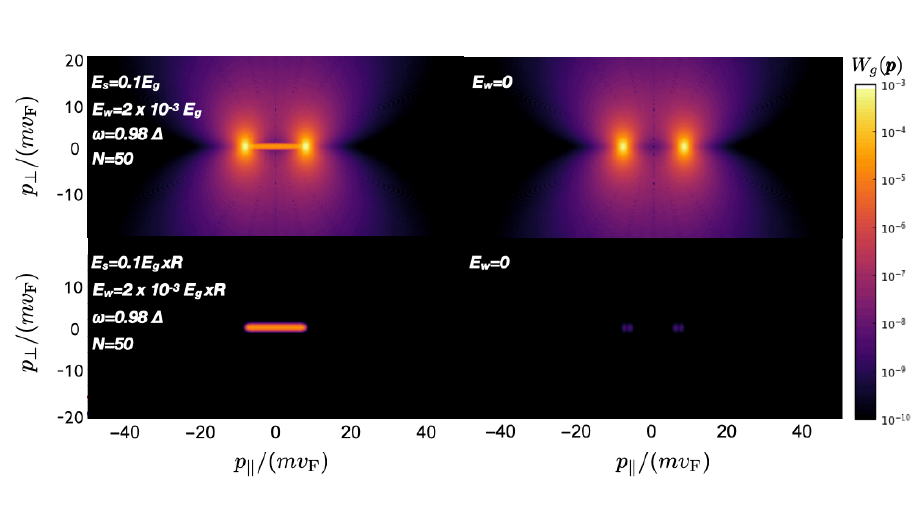}
\caption{Density plots showing the quasiparticle spectrum associated with a sharply edged field (upper panel).  The lower panel displays results linked to a field background with  smooth on/off sectors.  All pictures were generated by taking the number of cycles $N=50$ ($T = 30\;\rm ps$).  The remaining parameters have been taken as in Fig.~\ref{fig:1.1}. The color legends of both $W_{\mathpzc{g}}(\pmb{p})$ are given in logarithmic scales.}
\label{fig:1.3}
\end{figure*}

Further insight into the quasiparticle spectrum  can be inferred from the right panel of  Fig.~\ref{fig:2}. There,  the behavior of $W_{\mathpzc{g}}(p_\perp)$ with the momentum perpendicular to the external electric field at $p_\parallel=0$ is depicted for the case in which the transitions are driven by a sharply edged field (top) and one  in which the field switching occurs smoothly  (bottom).  The  picture  indicates  clearly  that the quasiparticles promoted to the conduction band are more likely to appear  either at rest or moving along the strong field direction. The described trend is reproduced by the approximated expression in Eq.~\eqref{distributionfunction}, the results of which deviate from the exact result as  $p_\perp$ increases.  Such a  behavior becomes noticeable when the external field is characterized by abrupt on/off edges. Indeed, in this scenario, the  solid curves obtained from solving the Boltzmann-Vlasov equation show  pronounced oscillatory tails for  $p_\perp>\mathpzc{mv}_{\mathrm{F}}$, which are absent from the outcomes linked to the case in which the field comprises smooth ramping and deramping phases.  We remark that  the width $\Delta p_\perp =2\mathpzc{mv}_{\mathrm{F}}$ can  be used to obtain a glimpse of the transversal extension $\Delta x_\perp \gtrsim \Delta p_\perp^{-1}$ of the wave packet. Indeed, by taking $\Delta=7\;\rm meV$, one finds $\Delta x_\perp \gtrsim  93\;\rm nm$.

The fact that the longitudinal and transversal widths $\Delta p_\parallel$ and $\Delta p_\perp$ differ from each other evidences that the quasiparticle spectrum has a remarkable anisotropic character, a fact that is verified in Fig.~\ref{fig:1.3}. The  behavior of the spectrum  for a field with sharp edges is depicted in the upper panel, while the counterpart with smooth ramping and deramping sectors is shown in the lower panel. The left density plots have been generated by setting the strong field strength to $E_s=0.1E_{\mathpzc{g}}$, whereas the remaining parameters coincide with those used for Fig.~\ref{fig:2}.  Here, the right pictures correspond to the case in which the fast oscillating wave is turned off [$E_w=0$].   In the upper panel, the  most intensely colored  regions show the characteristic peaks linked to the field given by Eq.~\eqref{EField}. The previously described oscillatory sector  corresponds to a less intense colored band that lies between the most bright zones (peaks) in the upper left  figure. This sector is, however, absent in the picture  on the right. A remarkable feature exhibited in the upper  panel of  Fig.~\ref{fig:1.1}  is that the areas surrounding the peaks with and without the fast oscillating mode are comparable.  This provides evidence that the difference between observables linked to the respective distribution functions could be much less sensitive to the finite-size effects of the field background.

Observables, such as the number of transitions per unit area from the valence to the initially empty conduction band, are directly influenced by the spectrum properties:
\begin{equation}
\begin{split}
\mathpzc{N}_{\;\mathpzc{g}}=g_sg_v \int_{\mathrm{BZ}} \frac{d^2p}{(2\pi)^2}\; W_{\mathpzc{g}}(\pmb{p}).
\label{Ninitial}
\end{split}
\end{equation}Here,   $g_s=2$ ($g_v=2$) accounts  for the spin (valley) degeneracy and the integration is limited by the Brillouin zone.  As we will see shortly, under particular circumstances, the observable above determines the long-term current excited by the external field configuration [see Eq.~\eqref{currentasymptotic}]. Conceptually,  $\mathpzc{N}_{\;\mathpzc{g}}$  is nothing else but the volume below the surfaces exhibited by any of the pictures in Fig.~\ref{fig:1.3}.  Observe that in the case driven by a sharp-edge field, the height of the peaks is roughly two orders of magnitude larger than the mean altitude of the region between them. As a consequence, one can anticipate a dominance of the side peaks  in the problem's phenomenology 
 unless the extent of the middle part is sufficiently large to mitigate their impact. In other words,  as the volume bounded by the oscillatory sector grows, the likelihood of detecting the effect will rise.  For a given strong field,  this condition can be achieved by increasing the operating time $T$  while the restriction in Eq.~\eqref{timerestrictions} is fulfilled.

\section{Residual field-induced current  \label{observabledcp}}

\subsection{Characterizing the Franz-Keldysh current in a semiconductor graphene monolayer\label{FKEACURR}}

The mean current density along the applied electric field direction, induced by the  creation of electron-hole pairs, splits into two contributions \cite{Gravilov2012,dora1}:
\begin{equation}
j(t)=j_{\mathrm{con}}(t)+j_{\mathrm{pol}}(t)
\label{jparallel}
\end{equation}  identified as  the conduction $j_{\mathrm{con}}(t)$ and polarization $j_{\mathrm{pol}}(t)$ currents \cite{kluger1991,Bloch1999,Otto2019}. Explicitly,  
\begin{equation}
\begin{split}
&j_{\mathrm{con}}(t)=2 e g_s g_v \int_{\mathrm{BZ}} \frac{d^2 p}{(2\pi)^2}\frac{\partial\mathpzc{w}_{\pmb{p}}(t)}{\partial p_\parallel}W_{\mathpzc{g}}(\pmb{p};t),\\
&j_{\mathrm{pol}}(t)=2 e g_s g_v \int_{\mathrm{BZ}}  \frac{d^2 p}{(2\pi)^2}\frac{\mathpzc{w}_{\pmb{p}}(t)}{eE(t)}\dot{W}_{\;\mathpzc{g}}(\pmb{p};t).
\end{split}\label{condandpolarcurrent}
\end{equation} The factor $2$  in these  expressions  is due to the contribution  of  both the quasiparticles and holes.   Even while the  formulas above  serve as the starting point for the following  numerical investigation, we are also driven to derive an analytical expression for the current density that will allow us to determine how it scales with various external field parameters. To this end, let us first consider the residual conduction current $j_{\mathrm{con}}\equiv\lim_{t\to T/2}j_{\mathrm{con}}(t)$. Observe that, in this scenario, if the condition $\frac{1}{2}eE_sT\gg \vert p_{\parallel,\perp}\vert$ is satisfied [see below Eq.~\eqref{distributionfunction}], the carriers' velocity saturates at the Fermi velocity, i.e., $\lim_{t\to T/2}\partial\mathpzc{w}_{\pmb{p}}(t)/\partial p_\parallel\to \mathpzc{v}_{\mathrm{F}}$. Under such  circumstances the residual  conduction  current  approximates  $j_{\mathrm{con}}\approx 2e\mathpzc{v}_{\mathrm{F}}\mathpzc{N}_{\;\mathpzc{g}}$, where   $\mathpzc{N}_{\;\mathpzc{g}}$ is  the density of electrons that occupies the initially empty upper Dirac cone [see Eq.~\eqref{Ninitial}]. Now, electric dipoles are created in conjunction with each electron-hole pair that is produced in the course of the transitions.  In  a semi-classical picture, this dipole moment becomes  separated by an instantaneous  barrier width  $\mathpzc{l}_{\rm eff}(t)=2\mathpzc{w}_{\pmb{p}}(t)/[eE(t)]$  and has a magnitude $\mathpzc{p}(t)=2e\mathpzc{l}_{\rm eff}(t)$. The polarization current is therefore caused by the rate at which these dipoles are created.  Observe that, at a time for which the electric field has been switched off,  the dipole is $\lim_{t\to T/2}\mathpzc{p}(t)=2eT\mathpzc{v}_{\mathrm{F}}$ and $j_{\mathrm{pol}}\equiv \lim_{t\to T/2} j_{\mathrm{pol}}(t)\approx e\mathpzc{v}_{\mathrm{F}}T\dot{\mathpzc{N}}_{\;\mathpzc{g}}$. It is worth remarking that   $\lim_{t\to T/2}\dot{W}_{\; \mathpzc{g}}(\pmb{p},t)\propto Q(\pmb{p},T/2)\sim T^{-2}$, provided the condition  $T\gg 2\vert p_{\perp,\parallel}\vert/(eE_s)$  applies  [see below Eq.~\eqref{distributionfunction}].  As a result, the residual polarization current   $j_{\mathrm{pol}}\propto T^{-1}$ is suppressed and can, thus, be safely ignored in our calculation. Consequently,  Eq.~\eqref{jparallel} approaches to 
\begin{equation}\label{currentasymptotic}
j\equiv j(T/2)\approx 2e\mathpzc{v}_{\mathrm{F}}\mathpzc{N}_{\;\mathpzc{g}}.
\end{equation}

An approximated expression for the current density  can be established  by  inserting   Eq.~\eqref{distributionfunction}  into Eq.~\eqref{Ninitial} and integrating $p_\parallel$ and $p_\perp$ out. 
To be consistent with  both the application range of the Dirac model [see discussion  below Eq.~(\ref{vlasovgraphene})] and the condition under which Eq.~\eqref{distributionfunction}  was derived,  the integration domains have to  be limited by  $\pm\frac{1}{2}eE_sT$. Nevertheless observe that,  from a practical prospect, the fast damping of the integrand in $p_{\perp}$  allows us to extend the corresponding integration limits to $\pm\infty$ without introducing an appreciable error.  Considering the periodicity  of $W_{\mathpzc{g}}(\pmb{p})$  in  $p_\parallel$, as well as its even character in both $p_\parallel$ and $p_\perp$ we end up with
\begin{widetext}\begin{equation}
\begin{split}
j&\approx 2e \mathpzc{v}_{\mathrm{F}}^{\nicefrac{1}{2}} T \frac{(eE_s)^{\nicefrac{3}{2}}}{\pi^{2}}e^{-\pi\frac{E_{\mathpzc{g}}}{E_s}} \left\{1+ \frac{\varepsilon^2\pi^2}{4}e^{\pi\frac{E_{\mathpzc{g}}}{E_s}} \int_{\frac{E_{\mathpzc{g}}}{E_s}}^\infty\frac{ds\; s^2}{\sqrt{s-\frac{E_\mathpzc{g}}{E_s}}}e^{-\pi s}\left\vert _1\tilde{F}_1\left(1-\frac{i}{2}s;2;\frac{i}{2}\gamma \frac{2\omega}{\Delta}\right)\right\vert^2\right\},
\end{split}\label{main}
\end{equation}\end{widetext}where the change  of variables  $s=\epsilon_\perp^2/(eE_s\mathpzc{v}_\mathrm{F})$ has  been carried out.  In a scenario where the graphene flake has a trivial bandgap [$\Delta=0$], the second term of the expression above is negligible as compared to its leading-order contribution $j_{\Delta=0}\approx 2e \mathpzc{v}_{\mathrm{F}}^{\nicefrac{1}{2}} T (eE_s)^{\nicefrac{3}{2}}/\pi^{2}$.  We remark that Eq.~\eqref{main} applies whenever $E_s\gg E_w$ and for $T\ll T_{\mathrm{Bloch}}$. As no further restriction is required on the strong electric field, this expression can be a priori utilized to explore the over-critical field regime where $E_s\gg E_{\mathpzc{g}}$. However, care has to be taken in such a situation because the circumstances are favorable for a large number of transitions, and the internal field created by electrons and holes might deplete considerably the external field strength.  This backreaction effect is, however, not taken into account, neither in the transport equation [see Eq.~\eqref{vlasovgraphene}] nor in the approximated formulas in Eqs.~\eqref{distributionfunction} and \eqref{main}.  Because of this limitation, our efforts are focused on investigating the subcritical regime of the strong field $E_s\lesssim E_{\mathpzc{g}}$. Still, if the strong field is close to the critical field,  Eq.~\eqref{main}  can be used for obtaining a glimpse of the Franz-Keldysh current.  Observe that, in the scenario characterized by the condition $\gamma\ll \Delta/\omega$, i.e. when $\omega/\Delta\ll (E_s/E_{\mathpzc{g}})^{1/2}$  the third argument of the regularized hypergeometric function is much smaller than  unity. In such a case,  $_1\tilde{F}_1\approx 1$ and  the current density  is mainly driven by  the tunneling from valence to conduction band:
\begin{equation}
\begin{split}
j\approx 2e \mathpzc{v}_{\mathrm{F}}^{\nicefrac{1}{2}}\frac{(eE_s)^{\nicefrac{3}{2}}}{\pi^2} e^{-\pi\frac{E_\mathpzc{g}}{E_s}}T\left\{1+\frac{\varepsilon^2\pi^2}{4}\left(\frac{E_{\mathpzc{g}}}{E_s}\right)^2\right.\\
\times\left.\left[1+\frac{1}{\pi}\frac{E_{s}}{E_{\mathpzc{g}}}+\frac{3}{4\pi^2}\left(\frac{E_{s}}{E_{\mathpzc{g}}}\right)^2\right]\right\}.
\end{split}
\label{Beyondschwingerrate}
\end{equation} In the particular case where $\varepsilon=0$,  this formula reduces to the Landau-Zener current \cite{dora1}:
\begin{equation}
\begin{split}
j_{\mathrm{LZ}}\approx 2e \mathpzc{v}_{\mathrm{F}}^{\nicefrac{1}{2}}\frac{(eE_s)^{\nicefrac{3}{2}}}{\pi^2} e^{-\pi\frac{E_\mathpzc{g}}{E_s}}T,
\end{split}
\label{LZJ}
\end{equation}
 which turns out to be the leading order contribution when $\varepsilon\neq0$  and $E_s<E_{\mathpzc{g}}$. We note that this will be the case whenever $\omega/\Delta\ll (E_s/E_{\mathpzc{g}})^{1/2}<1$ .
 
Now, when $\gamma\gg \Delta/\omega$, and $E_s/ E_{\mathpzc{g}}\ll 1$ the current  is induced by the absorption of  one  photon from the fast-oscillating wave [see Eq.~\eqref{largekeldych}]. In this case, 
 \begin{equation}\label{perturbativecurrent}
 j\approx e \xi_w^2 T \frac{\Delta^2 \omega}{16 \mathpzc{v}_{\mathrm{F}} }\left(1+\frac{\Delta^2}{\omega^2}\right)\Theta(\omega-\Delta),
 \end{equation}where  $\xi_w=eE_w/(\omega\mathpzc{mv}_{\mathrm{F}})\ll1$ is the laser intensity parameter, and $\Theta(x)$ is the unit-step function. We remark  that, unlike its three-dimensional counterpart, this current does not vanish at the pair production threshold  [see cyan curve in Fig~\ref{fig:7}]. Such a behavior--due to graphene's low-dimensionality--is completely analogous to the one found in the perturbative regime of the Breit-Wheeler process, where non-vanishing rates at the energy threshold arise when odd numbers of photons are absorbed \cite{Golub:2019vzb,Golub:2021nhj}.  The Breit-Wheeler process, however, manifests this unusual behavior in the third order in perturbation theory, which makes its experimental verification difficult.  Measuring the current in Eq.~\eqref{perturbativecurrent}  provides, therefore, an efficient way to confirm this low-dimensional effect. Indeed, in a scenario characterized by  $E_w=2\times 10^{-3}E_{\mathpzc{g}}$, $\Delta=2\; \rm meV$,  and an interacting time $T=T_{\mathrm{Bloch}}/3$ with $E_s\to E_w$  [see   Eq.~\eqref{Blochscale}], we found  $j\approx 2.4\times 10^{-8}\; \rm A/\mu m$. Hence, the current to be measured as a result of the rate's lifting at the threshold is $\mathcal{I}=j\ell>2.4\;\rm \mu A$, if a graphene sample with characteristic length $\ell>100\;\rm \mu m$ is used.
 
\subsection{Current density in the combined subthreshold and subcritical regimes}

\begin{figure*}
\centering
\includegraphics[width=0.48\textwidth]{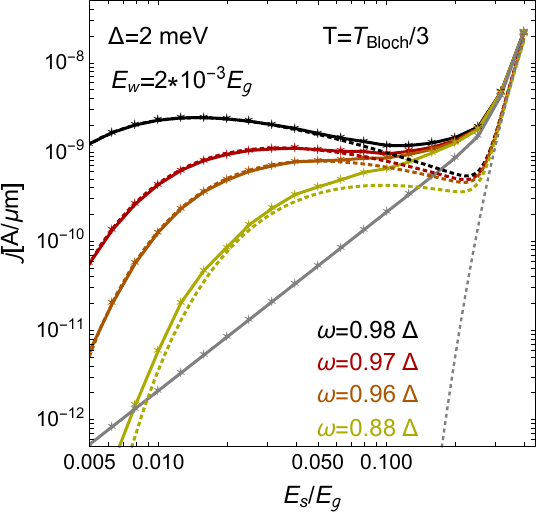}
\includegraphics[width=0.48\textwidth]{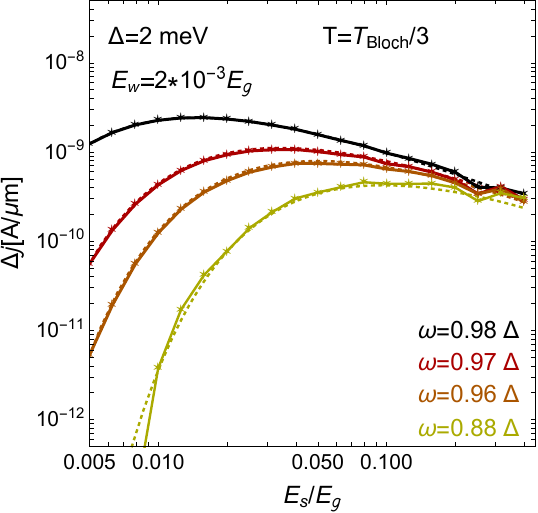}
\caption{Current density  and its difference with respect to the case driven by the strong field solely  at the stage in which the  field  is turned off.  In both panels, the plot markers show the numerical results, which  have been obtained by adjusting  the interacting time to one-third of the corresponding Bloch's scale, i.e., the time is not fixed along a colored curve but rather varies from one marker to another. For comparisons, the outcomes resulting from the analytical formula  \eqref{main}  have been included [dashed curves]. In each panel, curves sharing the same color have been obtained by using the same parameters.  Particularly, the  curves in gray correspond to  $E_w=0$.  All curves have been generated by  setting  a  gap  $\Delta=2\;\rm meV$ leading to a critical field $E_\mathpzc{g}=\Delta^2/(4e\mathpzc{v}_{\mathrm{F}})\approx 15.6\ \rm V/cm$.}
\label{fig:4}
\end{figure*}

This section will focus on the subcritical field regime [$E_s\ll E_{\mathpzc{g}}$]  where the Franz-Keldysh current combines the tunneling effect with the absorption of one photon having energy just below the bandgap [$\omega \lesssim\Delta$].
We will begin our analysis by first considering an electric field with sharp edges, as provided by Eq.~\eqref{EField}.  In order to assess the enhancement induced by the fast-oscillating wave as compared to the case in which it is turned off  [solid gray style],  we show in  Fig.~\ref{fig:4} [solid  curves]  the  dependence of $j$ and  $\Delta j\equiv j(T/2)-j_{\mathrm{LZ}}(T/2)$  with respect to the strong electric field strength $E_s$ within the  range  $E_s\in[4\times 10^{-3},4\times 10^{-1}]E_{\mathpzc{g}}$.   We have chosen a gap  $\Delta=2\;\rm meV$  as reference value for our assessment, leading to the characteristic  field scale  $E_{\mathpzc{g}}=\Delta^2/(4e\mathpzc{v}_{\mathrm{F}})\approx 15.6\; \rm V/cm$.   To optimize the enhancement and minimize the impact of the finite-size effect of the external field, the interacting time has been adjusted for each field strength in such a way that it accounts for one third of the characteristic Bloch's scale. This means for instance  that,  at $\omega=0.98\;\Delta$ and $E_s=0.004E_{\mathpzc{g}}$,  the interacting time $T\approx1.7\times 10^2\; \rm ns$, whereas  $T\approx 1.7\; \rm ns$ for  $E_s=0.4E_{\mathpzc{g}}$.   In both panels, the results  have been  obtained by setting the weak field strength to $E_w=2\times 10^{-3}\;E_{\mathpzc{g}}$. In the course of the numerical computations, we have noted that the polarization current in the domain under consideration is insignificant compared to the corresponding conduction current.  Each solid curve there is linked to a particular laser frequency, and as this parameter increases, it becomes clear that the results differ significantly from the ones associated with the standard Landau-Zener transitions [solid gray style].  This enhancement is attributed to the absorption of photons from the fast-oscillating wave in the presence of the strong field, and it becomes particularly noticeable in the range of $E_s$  exhibited in Fig.~\ref{fig:4}.  It confirms  that there are two paths to boost the interband transitions of electrons in a photo-catalyzed system. Namely, by strengthening the strong field or through absorption of  quanta from the fast-oscillating  beam. The former channel rules the process as $E_s$ increases,  while the latter dominates as the contrary condition occurs.  

Manifestly, the left panel in Fig.~\ref{fig:4} exhibits a nonlinear dependence of the current $j$ on the strong field $E_s$. According to Ohm's law [$j=\sigma E_s$], this implies a field-dependent conductivity $\sigma$ which can be  read off from Eq.~\eqref{main}. In the left panel of Fig.~\ref{fig:4}  the outcome resulting from the analytical formula [see Eq.~\eqref{main}] has been included in dashed style.  These curves, however, differ from the numerical ones in some portions along the electric field sector that is displayed.  The situation is more pronounced in the case where the weak field is turned off [compare the solid and dashed gray curves], where the result deviates substantially from the exponential damping linked to the standard Landau-Zener current  [see Eq.~\eqref{LZJ}]. This mismatch is due to the strong field's sharp edges, which practically control the phenomenology  via the side peaks that arise in $W_{\mathpzc{g}}$ [see Fig.~\ref{fig:1.1}, upper panel].  Although less severe, the impact of these maxima is also visible in  the assisted scenario, particularly for field strengths $E_s>5\times 10^{-2}E_{\mathpzc{g}}$. In order to mitigate their  influence, it is feasible to measure the currents with and without the fast-oscillating field separately, as is done in modulation spectroscopy,  a technique traditionally employed  for measuring the Franz-Keldysh effect  \cite{Cardona}.  The difference between these currents $\Delta j\equiv j(T/2)-j_{\mathrm{LZ}}(T/2)$  constitutes a testable indicator for the dynamically assisted Landau-Zener transitions, and its behavior is shown in the right panel of  Fig.~\ref{fig:4}.  Because the effects of the peaks in $W_{\mathpzc{g}}(\pmb{p})$ are practically cancelled out, it is visible that there is good agreement between the numerical and analytical results.

For a frequency  $\omega\approx0.98\;\Delta$, the black curves in both pictures tend to exhibit a local maximum  at $E_s\approx 1.5 \times 10^{-2} E_{\mathpzc{g}}$.  Indeed, for this  parameter combination the  current amounts to  $j_{\mathrm{max}}\approx\Delta j_{\mathrm{max}}\approx 2.4 \times 10^{-9}\; \rm A/\mu m$, which improves  the unassisted scenario by three orders of magnitude approximately. The emergence of this local peak  is a remarkable feature that would allow us  to  verify the process with  a  field strength $E_s$ which is two orders of magnitude weaker than  $E_{\mathpzc{g}}$ and almost an order of magnitude larger than the weak field strength.   We remark that if the previous conditions are met experimentally, the current  to be measured in samples with characteristic length $\ell=100\;\mu\rm m$ will be $\mathcal{I}=j_{\mathrm{max}}\ell\gtrsim 0.24\; \mu\rm A$  at the stage in which the assisted field configuration is turned off. Noteworthy, this   prediction  could be  measurable with presently available technology.  Indeed,  in  graphene layers with gapless band structures  [$\Delta=0$], photo-excited currents  $\sim\rm pA$ have already been detected via Landau-Zener-St\"uckelberg  interferometry \cite{Takuya,Heide}. It is worth remarking that, since there is a good agreement between the analytical and numerical results at the maximum, the change in slope from positive to negative around this field strength can be directly related to the number of produced quasiparticles [see Eq.~\eqref{currentasymptotic}]. The aforementioned peak has been numerically  verified  to remain essentially unchanged when  a carrier envelope phase $\phi_{\mathrm{CEP}}$ is inserted into the fast-oscillating wave, and it is varied continuously within $[0,\pi]$.  Moreover, a comparison between the black and red curves in  Fig.~\ref{fig:4}  reveals that the location of the maximum varies with changes in frequency and strong field strength. Indeed, for $\omega\approx0.97\;\Delta$, the maximum current is reached at $E_s\approx 0.04E_{\mathpzc{g}}$. 
 \begin{figure}
\includegraphics[width=0.48\textwidth]{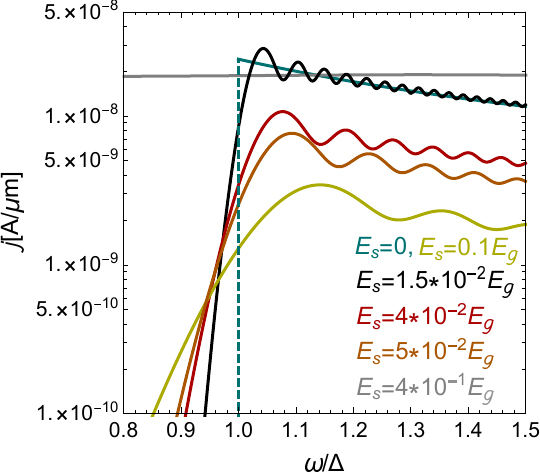}
\caption{Dependence of the residual current density on the frequency of the fast-oscillating wave. The curves colored in black, red, orange yellow and gray are based on the analytical formula given in  Eq.~\eqref{main}, whereas the one in cyan corresponds to Eq.~\eqref{perturbativecurrent}. The absorptive threshold is represented by the  dashed line for reference. The  benchmark parameters and notation used in Fig.~\ref{fig:4} have also been adopted here.}
\label{fig:7}
\end{figure}
We remark that for the strong fields at which the maxima arise, the Landau-Zener contribution in Eq.~\eqref{main} is exponentially suppressed  compared to its second contribution, i.e.
\begin{equation}
\begin{split}
j\approx\Delta j\approx  \frac{1}{2}e \mathpzc{v}_{\mathrm{F}}^{\nicefrac{1}{2}} T \frac{(eE_w)^2}{\sqrt{eE_s}} \int_{\frac{E_{\mathpzc{g}}}{E_s}}^\infty\frac{ds\; s^2}{\sqrt{s-\frac{E_\mathpzc{g}}{E_s}}}e^{-\pi s}\\
\times\left\vert _1\tilde{F}_1\left(1-\frac{i}{2}s;2;\frac{i}{2}\gamma \frac{2\omega}{\Delta}\right)\right\vert^2
\end{split} \label{main2}
\end{equation} with $E_w\ll E_s\ll E_{\mathpzc{g}}$. This is why the left sectors of each picture in Fig.~\ref{fig:4}  coincide quantitatively and qualitatively. According to this analysis, the locations of the maximum are unaffected by the amplitude $E_w$ of the fast oscillating wave, but their heights scale quadratically with it.

Fig.~\ref{fig:7}  provides a further  overview of the peaks' rise  with the fast-oscillating mode's frequency $\omega$. The picture covers an interval, ranging from the sub-threshold domain  $\omega<\Delta$ to the over-threshold regime $\omega\geqslant\Delta$, and includes the result associated with Eq.~\eqref{perturbativecurrent} (cyan solid  curve). The current density linked to this case  exhibits the jump at the threshold $\omega=\Delta$ (cyan dashed line) discussed previously in Sec.~\ref{FKEACURR}. By setting the strong field at the values for which the maxima of the black, red, and orange curves in the left panel of Fig.~\ref{fig:4} are hit, respectively, the results colored in black, red, and orange in Fig.~\ref{fig:7}  have been generated. Conversely, the gray and yellow curves were obtained by setting $E_s=0.4E_{\mathpzc{g}}$ and $E_s=0.1E_{\mathpzc{g}}$, respectively. We observe that, below the threshold $\omega<\Delta$, the curves tend to stick closer to  the vertical line as both $E_s\to0$ and $\omega\to \Delta$. As this occurs, the absorption channel becomes more and more efficient in inducing electron transitions from the valence to the conduction band, and the current density tends to saturate to the nontrivial perturbative value $j\approx 2.4\times 10^{-8}\; \rm A/\mu m$  dictated by Eq.~\eqref{perturbativecurrent}. For frequencies above the threshold $\omega\geqslant\Delta$,  $j$ exhibits the characteristic oscillatory behavior of the Franz-Keldysh effect. Indeed, Fig.~\ref{fig:7}  reveals the hallmarks of this process in graphene by illustrating how the current solely associated with the perturbative absorption channel varies when a nonperturbative electric field is turned on.

\begin{figure}
\includegraphics[width=0.45\textwidth]{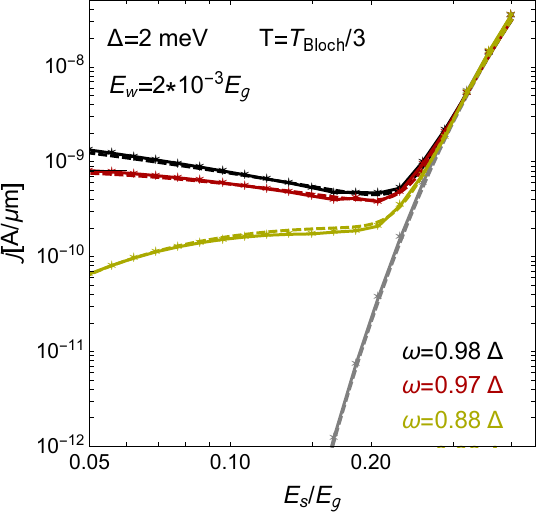}
\caption{Dependence of the current density on a strong field with smooth turn-on/off sectors.  The  benchmark parameters and notation used in Fig.~\ref{fig:4} have also been adopted here. }
\label{fig:5}
\end{figure}

Coming back to the combined  sub-threshold and subcritical  regime, Fig.~\ref{fig:5}  is intended to provide insight about the trend of the current density with the change of $E_s$ in the case where the external background is characterized by smooth turning on and off phases. The picture shows the $E_s$ region where the numerical  findings related to abrupt ramping and deramping functions differ from the corresponding analytical results. In contrast to the left panel in Fig.~\ref{fig:4}, the agreement is noticeably good in both the assisted and unassisted scenarios, the latter one [dotted gray curve] exhibiting the known exponential falling for small field strength. The fact that the Landau-Zener trend is less sensitive in the present situation to  the field's finite size can be understood as a direct consequence of the chosen interaction time, which makes the resonances linked to the side-band terms negligible as we pointed out in Sec.~\ref{SPDFasymptotic}.  We note that, compared to the situation dealing with abrupt turn-on/off phases, the enhancement caused by the assisting fast-oscillating wave is increased by several orders of magnitude in the current field configuration. For instance,  at $E_s=0.1 E_{\mathpzc{g}}$  and $\omega=0.98\Delta$ the improvement with smooth ramping and de- ramping phases with $\sin^2[\frac{1}{2}\pi(1-6t/T)]$-shape exceeds  the one exhibited in   left panel of Fig.~\ref{fig:4}  by a factor $\sim10^8$. This is unambiguous proof that measurements of the induced current in this situation are less sensitive to the effects of the field's finite size.

\section{Conclusions} \label{sec:numerical}

Summarizing, we have investigated the Franz-Keldysh effect in semiconductor graphene.  Our study, based on the Dirac-like model, indicates that the electronic transitions are strongly influenced by the turn-on/off stages of the field background, being less affected when these phases occur smoothly on time scales larger than the formation time needed to create an electron-hole pair.  A general expression for the associated current density was derived, and its various asymptotic regimes of field strength and frequency have been explored.   In the scenario in which the strong field effect is suppressed as compared to the one induced by the fast-oscillating wave, the current shows the characteristic threshold jump that other perturbative absorptive channels manifest when a band gap is induced in  graphene. It has been argued that measurements made using the proposed setup can benefit the detection of this anomalous behavior rooted in the low dimensionality. We have verified that in the combined subcritical field and subthreshold regimes, where the absorption process competes against the tunneling effect induced by the strong field, the interband transitions of electrons are highly stimulated as compared to the prediction linked to the standard Landau-Zener effect.  We have established further conditions that would ensure process optimization  and pointed out that measurements of the generated current can verify  the solid-state analog of the dynamically-assisted Schwinger mechanism in QED.

\section*{Acknowledgments}

RE acknowledges funding by the Deutsche Forschungsgemeinschaft (DFG, German Research Foundation), under Projektnummer 277101999 -- TRR 183 (project B04), 
and under Germany's Excellence Strategy -- Cluster of Excellence Matter and Light for  Quantum Computing (ML4Q) EXC 2004/1 -- 390534769.  The authors thank an anonymous referee for drawing their attention to  important aspects that significantly improved the manuscript.

\end{document}